Relative abundances of $CO_2$, CO, and $CH_4$ in atmospheres of Earth-like lifeless planets

Yasuto Watanabe[12*†], Kazumi Ozaki[3456*†]

## Abstract

Carbon is an essential element for life on Earth, and the relative abundances of major carbon species ($CO_2$, CO, and $CH_4$) in the atmosphere exert fundamental controls on planetary climate and biogeochemistry. Here, we employed a theoretical model of atmospheric chemistry to investigate diversity in the atmospheric abundances of $CO_2$, CO, and $CH_4$ on Earth-like lifeless planets orbiting Sun-like (F-, G-, and K-type) stars. We focused on the conditions for the formation of a CO-rich atmosphere, which would be favorable for the origin of life. Results demonstrated that elevated atmospheric $CO_2$ levels trigger photochemical instability of the CO budget in the atmosphere (i.e., CO runaway) owing to enhanced $CO_2$ photolysis relative to $H_2O$ photolysis. Higher volcanic outgassing fluxes of reduced C (CO and $CH_4$) also tend to initiate CO runaway. Our systematic examinations revealed that anoxic atmospheres of Earth-like lifeless planets could be classified in the phase space of $CH_4/CO_2$ versus $CO/CO_2$, where a distinct gap in atmospheric carbon chemistry is expected to be observed. Our findings indicate that the gap structure is a general feature of Earth-like lifeless planets with reducing atmospheres orbiting Sun-like (F-, G-, and K-type) stars.

[1] Department of Earth and Planetary Science, The University of Tokyo, Hongo 7-3-1, Bunkyo-ku, Tokyo, Japan. 113-0033

[2] Meteorological Research Institute, Japan Meteorological Agency, Nagamine 1-1, Tsukuba, Ibaraki, Japan. 305-0052

[3] Department of Earth and Planetary Sciences, Tokyo Institute of Technology, Ookayama 2-12-1, Meguro-ku, Tokyo, Japan. 152-8551

[4] Earth-Life Science Institute, Tokyo Institute of Technology, Ookayama 2-12-1, Meguro-ku, Tokyo, Japan. 152-8551

[5] Nexus for Exoplanet System Science (NExSS), National Aeronautics and Space Administration, Washington, D.C. 20546, USA

[6] Alternative Earths Team, Interdisciplinary Consortia for Astrobiology Research, National Aeronautics and Space Administration, Riverside, CA 92521, USA

[*] Corresponding Author yasuto.watanabe.wess@gmail.com , ozaki.k.ai@m.titech.ac.jp

[†] These authors contributed equally to this work.

## 1. INTRODUCTION

The existence of over 5500 exoplanets has been confirmed since the discovery of the first in 1995 (Mayor and Queloz 1995). The focus of this field of research is gradually shifting from exoplanet detection to identification of signs of habitability and potential biosignatures. The traditional concept of planetary habitability has centered around the presence of liquid water on the planetary surface, given its vital role in supporting life on Earth (Hart 1979; Kasting et al. 1993, 2014; Selsis et al. 2007; Bartik et al. 2011; Seager et al. 2012; Kasting 2014). This concept is encapsulated in the habitable zone (HZ), which defines the circumstellar region in which a terrestrial planet with a $CO_2$–$H_2O$–$N_2$ atmosphere could sustain liquid water on its surface (Huang 1959; Hart 1979; Kasting et al. 1993; Williams and Kasting 1997; Underwood et al. 2003; Selsis et al. 2007; Spiegel et al. 2008; Abe et al. 2011; Kaltenegger and Sasselov 2011; Kopparapu et al. 2013, 2014). Dozens of exoplanets have been discovered within the HZ (Anglada-Escudé et al. 2016; Gillon et al. 2017; Meadows et al. 2018; Hill et al. 2023). However, the presence of an exoplanet in the HZ does not guarantee its habitability because a range of additional factors, such as atmospheric composition, climatic conditions, and the availability of bioessential elements, also play key roles in planetary habitability.

Carbon is an element essential for life on Earth, and it exerts fundamental controls on planetary climate and biogeochemistry. Both $CO_2$ and $CH_4$ are potent greenhouse gasses that exert major control on the global climate. Conversely, CO is crucial for the early evolution of life, serving as an important source of carbon and energy for microorganisms owing to its high thermodynamic free energy or low electrode potential (Ragsdale 2004; Kharecha et al. 2005; Seager et al. 2012; Kasting 2014; Catling and Kasting 2017). Consequently, detailed understanding of those factors that govern the relative abundances of $CO_2$, CO, and $CH_4$ in planetary atmospheres has far-reaching implications in the search for habitable planets beyond our solar system.

Earth's early atmosphere before the emergence of life is thought to have been predominantly composed of $CO_2$ (and $N_2$) (Miller and Urey 1959; Walker et al. 1983; Sagan and Chyba 1997). However, early theoretical studies of atmospheric photochemistry have recognized the possible existence of a state called CO runaway (Kasting et al. 1983; Zahnle 1986; Kasting 2014; Hu et al. 2020; Ranjan et al. 2022, 2023), in which the photochemical budget of CO is out of balance in the atmosphere. Under anoxic conditions, the primary production pathway for CO is photodissociation of $CO_2$ by UV radiation ($CO_2 + h\nu$ ($\lambda < 204$ nm) $\rightarrow$ CO + O). The rate of the reverse reaction is slow because of the spin-forbidden condition (Kasting et al. 1983; Zahnle 1986; Kasting 2014; Wogan et al. 2020; Ranjan et al. 2022), and therefore the primary removal of CO is performed via reaction with OH radicals (CO + OH $\rightarrow$ $CO_2$ + H), which are derived primarily from photodissociation of water ($H_2O + h\nu$ ($\lambda < 240$ nm) $\rightarrow$ H + OH). Consequently, under conditions where the rate of CO production increases to a level comparable to or larger than the rate of water vapor photolysis, the photochemical CO budget is imbalanced, which leads to CO accumulation over time until it becomes limited by surface deposition (Kasting et al. 1983; Kasting 2014; Ranjan et al. 2022).

The possibility of CO runaway is critical in resolving the fundamental problem regarding the origin of life on Earth because various organic compounds suitable for the prebiotic chemistry are more likely to form in a CO-rich atmosphere than in a $CO_2$-rich atmosphere (Bar-Nun and Chang 1983; McGlynn et al. 2020; Zang et al. 2022). Previous UV irradiation experiments demonstrated that a range of organic matter (e.g., aldehydes and organic acids) can be formed in an anoxic atmosphere containing CO, $N_2O$, and liquid water (Zang et al. 2022). Notably, CO might be important in the formation of peptides (Huber and Wächtershäuser 1997, 1998). Furthermore, recent findings regarding extremely negative carbon isotopic values in the modern Martian atmospheric CO (Alday et al. 2023; Aoki et al. 2023) and organic matter buried in the Gale crater (House et al. 2022) suggest the importance of $CO_2$ photodissociation—a major CO source—in the modern and past Martian atmosphere (Schmidt et al. 2013; Ueno et al. 2022; House et al. 2022; Yoshida et al. 2023).

Here, we employed a theoretical model of atmospheric chemistry to explore diversity in the abundances of $CO_2$, CO, and $CH_4$ in planetary atmospheres of Earth-like lifeless planets orbiting Sun-like (F-, G-, and K-type) stars to gain insights into the search for habitable planets that might facilitate the emergence of life.

## 2. METHODS

### 2.1. One-dimensional photochemical model

We employed a vertically resolved one-dimensional photochemical model, *Atmos* (Arney et al. 2016, 2018), which was originally developed by Jim Kasting and his colleagues (Kasting et al. 1983; Kasting 1990; Pavlov et al. 2001; Kharecha et al. 2005). The model divides a planetary atmosphere below the altitude of 100 km into 200 layers. In each atmospheric box, the chemical budgets of 58 long-lived species, 4 atmospheric particles, 11 short-lived species, and 1 inert species ($N_2$) are evaluated. The long-lived species are transported vertically within the atmosphere via eddy and molecular diffusion. The model considers the rates of production and loss of these species via 383 chemical and photochemical reactions. The effects of lighting, irreversible escape to space, and surface removal via rainout and dry deposition are also considered. The chemical reactions and reaction coefficients are adopted from the latest version of the *Atmos* model (Wogan et al. 2020; Watanabe et al. 2023). A two-stream approximation is used for the radiation transfer in the atmosphere (Toon et al. 1989). The absorption cross sections of $H_2O$ and $CO_2$ are updated following previous studies (Lincowski et al. 2018; Ranjan et al. 2020). The vertical profiles of the air temperature, number density of the air, and eddy diffusion coefficient used in our reference experiment are the same as those used in previous studies (Pavlov et al. 2001; Watanabe et al. 2023) (black line in Figure A1). The vertical mixing ratio of $H_2O$ is calculated based on local temperature. To determine how the prescribed temperature affects the model results, we conducted a sensitivity experiment with respect to the temperature profile (Section 3.4.). The mixing ratio of $N_2$ is maintained at 0.8 at any altitude. Surface removal of long-lived species by dry deposition is calculated using the deposition velocities adopted in a previous study (Watanabe et al. 2023). The concentration of $CO_2$ at the bottom of the atmosphere is treated as one of the model sensitivity parameters. For other atmospheric species, the lower and upper boundary conditions are given by the flux.

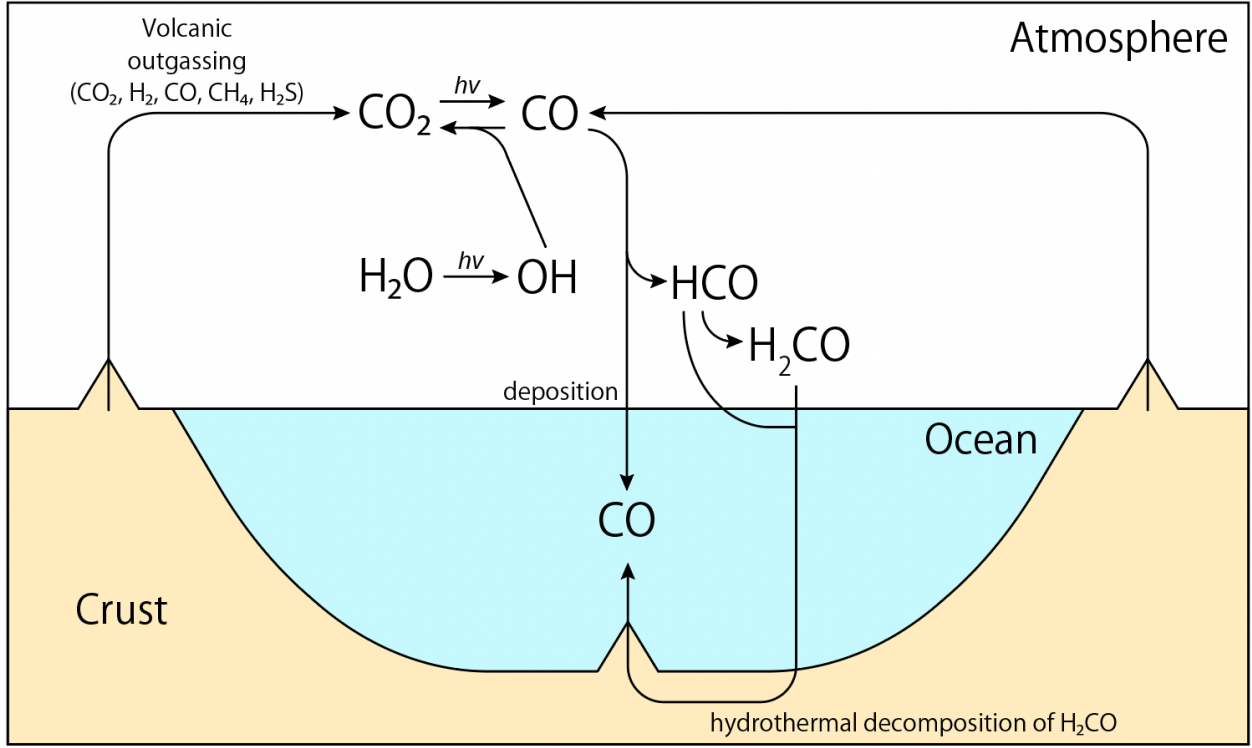

**Figure 1.** Schematic of the CO cycling in the ocean–atmosphere system considered in this study.

The total volcanic outgassing rate of reducing gases (in terms of $H_2$ equivalent) is expressed as follows:

$$\Phi_\uparrow(red) = \Phi_\uparrow(H_2) + \Phi_\uparrow(CO) + 4\Phi_\uparrow(CH_4) + 3\Phi_\uparrow(H_2S), \quad (1)$$

where $\Phi_\uparrow(X)$ represents the input flux of X to the atmosphere. The outgassing flux of both CO and $CH_4$ is assumed proportional to that of $H_2$ and determined as follows:

$$\Phi_\uparrow(CO) = \frac{1}{11}\Phi_\uparrow(H_2), \qquad \Phi_\uparrow(CH_4) = \frac{1}{27.5}\Phi_\uparrow(H_2) \quad . \qquad (2)$$

The factors in this equation are determined based on estimates of the outgassing rates of $H_2$, CO, and $CH_4$ from volcanoes and hydrothermal systems for the present Earth (Catling and Kasting 2017). It has been shown that these factors strongly depend on oxygen fugacity in magma (Wogan et al. 2020). For this reason, we assess the effects of the relative outgassing fluxes of CO and $CH_4$ on the results in Appendix B. The outgassing fluxes of $SO_2$ and $H_2S$ are assumed constant (i.e., 0.945 and 0.0945 Tmol S $yr^{-1}$, respectively) (Wogan et al. 2020; Watanabe et al. 2023).

The surface pressure, $P_S$, is calculated as follows:

$$P_s = max\left(P_0, \frac{pN_2}{1-f_{CO2}}\right), \qquad (3)$$

where $P_0$ represents the reference value of the total pressure (1.0 bar), $pN_2$ denotes the partial pressure of $N_2$ at the surface (0.8 bar), and $f_{CO2}$ denotes the mixing ratio of $CO_2$ at the surface. The partial pressure of $CO_2$ at the bottom of the atmosphere, $pCO_2$, can be written as follows:

$$pCO_2 = max\left(f_{CO2}P_0, \frac{f_{CO2}}{1-f_{CO2}}pN_2\right) \quad . \qquad (4)$$

2.2. CO cycle in the ocean–atmosphere system

The CO budget in the atmosphere can be written as follows:

$$\Phi_\uparrow(CO) + \Phi_{hyd}(CO) + \Sigma\Phi_{reac}(CO) + \Phi_{light}(CO) = \Phi_{\downarrow,dry}(CO), (5)$$

where $\Phi_\uparrow(CO)$ represents the volcanic outgassing flux of CO, $\Phi_{hyd}(CO)$ denotes the CO flux from the ocean to the atmosphere originating from the decomposition of $H_2CO$ and HCO at hydrothermal vents (see below), $\Sigma\Phi_{reac}(CO)$ is the net production rate of CO via chemical and photochemical reactions in the atmosphere, $\Phi_{light}(CO)$ is the CO production rate by lightning in the troposphere, and $\Phi_{\downarrow,dry}(CO)$ is the dry deposition rate of CO (Figure 1). Note that the production rate of CO in the atmosphere higher than the upper boundary of the model atmosphere is included in $\Sigma\Phi_{reac}(CO)$. Such high-altitude CO production may have a strong influence on the conditions of CO runaway, especially in the case of M-type stars (Hu et al. 2020; Ranjan et al. 2023), but does not strongly affect the conditions for CO runaway in the case of the Sun-like stars investigated in this study.

In the *Atmos* model, CO is removed from the lower boundary of the atmosphere via dry and wet deposition, whereas earlier studies considered only dry deposition (Kharecha et al. 2005; Kasting 2014). To represent the slow removal of dissolved CO in seawater, a low depositional velocity of $10^{-8}$ cm s$^{-1}$ is adopted, as in previous studies (Kharecha et al. 2005; Kasting 2014). The CO removed via wet deposition is assumed to immediately reach solubility equilibrium with the atmosphere. Therefore, the ultimate removal of CO at the bottom of the atmosphere is via slow abiotic chemical reactions (Kharecha et al. 2005; Kasting 2014).

In an anoxic atmosphere, formaldehyde ($H_2CO$) is effectively produced and deposited into the ocean. Hydrothermal decomposition of $H_2CO$ in the ocean has been proposed as a possible source of atmospheric CO (Holland 1984; Cleaves 2008; Watanabe et al. 2023):

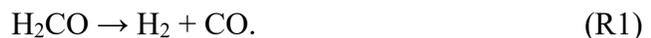

$$H_2CO \rightarrow H_2 + CO. \qquad (R1)$$

The depositional flux of HCO could also be large, and we consider the following possible decomposition pathway:

$$2HCO \rightarrow H_2 + 2CO. \tag{R2}$$

It is assumed that the deposited $H_2CO$ and HCO are readily decomposed by the above reactions and that $H_2$ and CO return to the atmosphere:

$$\Phi_{hyd}(H_2) = \Phi_\downarrow(H_2CO) + 0.5\Phi_\downarrow(HCO), \tag{6}$$

$$\Phi_{hyd}(CO) = \Phi_\downarrow(H_2CO) + \Phi_\downarrow(HCO). \tag{7}$$

### 2.3. Experimental setup

We conducted a series of numerical experiments with four different types of stellar spectra, assuming the young Sun (4.0 Ga) and F-/G-/K-type stars. For the case of the G-type star, the present solar spectrum (G2V) was adopted. For the F- and K-type stars, the spectra of σ Bootis (F2V) and ε Eridani (K2V) were employed, respectively (Figure 2). These stellar spectra were scaled such that the received energy is equal to the present Earth condition, which allows exploration of the typical conditions of the orbital semi-major axis within the HZ of each central star (Arney et al. 2017; Schwieterman et al. 2019). For each stellar spectrum, sensitivity experiments with respect to the partial pressure of $CO_2$ at the bottom of the atmosphere, $pCO_2$, and the volcanic outgassing rate of reducing gasses, $\Phi_\uparrow(red)$, were performed. For improved convergence, each experiment was repeated three times using the results of the previous run as the initial conditions.

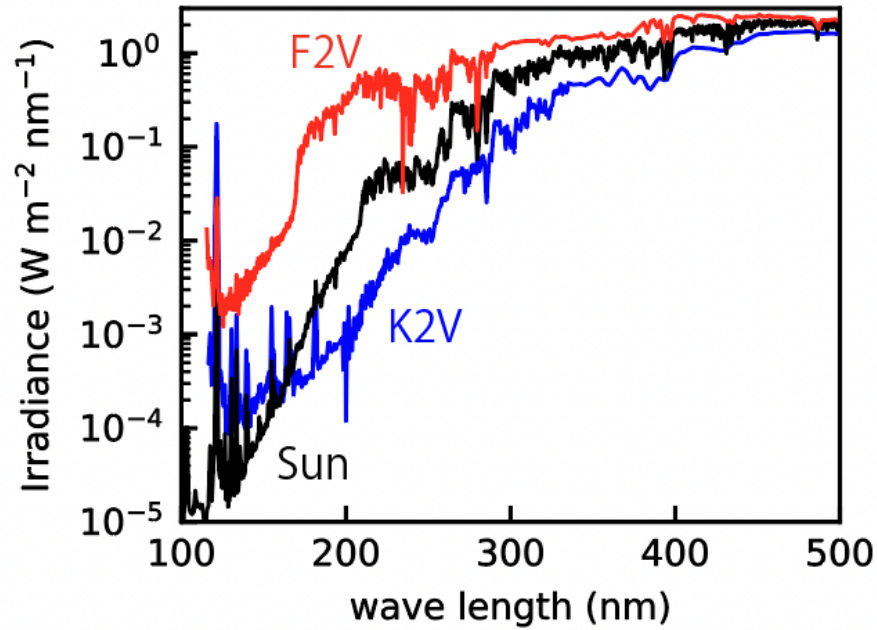

**Figure 2.** Spectra of Sun-like stars adopted in this study: Sun (G2V) (Thuillier et al. 2004), F-, and K-type stars (Segura et al. 2003).

## 3. RESULTS

3.1. Sensitivity to atmospheric $CO_2$ level and volcanic outgassing rate of reducing gasses

The underlying factors controlling CO runaway are (i) the atmospheric $CO_2$ level ($pCO_2$), (ii) the external input flux of reducing gases ($\Phi_{\uparrow}(red)$) that consume OH radicals, (iii) the availability of $H_2O$ (as a function of temperature), and (iv) the stellar spectrum of the central star (especially the irradiance in UV wavelengths). Here, we examine the response of atmospheric chemistry to changes in atmospheric $pCO_2$ and $\Phi_{\uparrow}(red)$ assuming the spectrum of the young Sun (4 Ga).

In the $pCO_2$ experiment, we varied atmospheric $pCO_2$ level from $5 \times 10^{-4}$ to ~2 bar while maintaining a constant $\Phi_{\uparrow}(red)$ (~46.2 Tmol $H_2$ equiv. $yr^{-1}$). The results demonstrate that the atmospheric partial pressure of CO, $pCO$, increases with increasing $pCO_2$ (Figure 3a). This can be attributed to enhanced $CO_2$ photodissociation (the primary source of CO) and suppressed availability of OH radicals (the primary sink of CO) linked to the shielding of tropospheric water vapor from UV radiation by $CO_2$ (see Appendix C for further explanation). Under low $pCO_2$ conditions (<0.2 bar), the external input of CO to the atmosphere via volcanoes and hydrothermal decomposition of $H_2CO$ and HCO is balanced mainly by photochemical removal (Figure 3b). We find that further increase in $pCO_2$ leads to marked increase in $pCO$. This corresponds to CO runaway, whereby photochemical CO removal is overwhelmed by source fluxes, leading to buildup of CO in the atmosphere until it is limited by surface deposition (Figure 3b). Indeed, under CO runaway conditions, dry deposition is one of the primary pathways of CO removal because the sink of the photochemical reactions becomes less effective and the production of CO by lightning is also enhanced. Production of CO from hydrothermal decomposition of $H_2CO$ and HCO is less effective under CO runaway conditions.

We also performed a sensitivity experiment with respect to $\Phi_\uparrow(red)$, which varied from ~1 to ~300 Tmol $H_2$ equiv. $yr^{-1}$ while maintaining constant atmospheric $CO_2$ (0.2 bar) (Figure 3d–3f). As might be expected, the abundance of reducing gases ($H_2$, CO, and $CH_4$) in the atmosphere increases with increasing $\Phi_\uparrow(red)$ (Figure 3d). Because these gases consume OH radicals, increase in $\Phi_\uparrow(red)$ leads to CO runaway (>~50–90 Tmol $H_2$ equiv. $yr^{-1}$). As in the $p$CO$_2$ experiment, the external input of CO is balanced by photochemical reactions in the atmosphere before CO runaway. However, once CO runaway occurs, photochemical reactions become a net source of CO and excess CO is removed via dry deposition (Figure 3e).

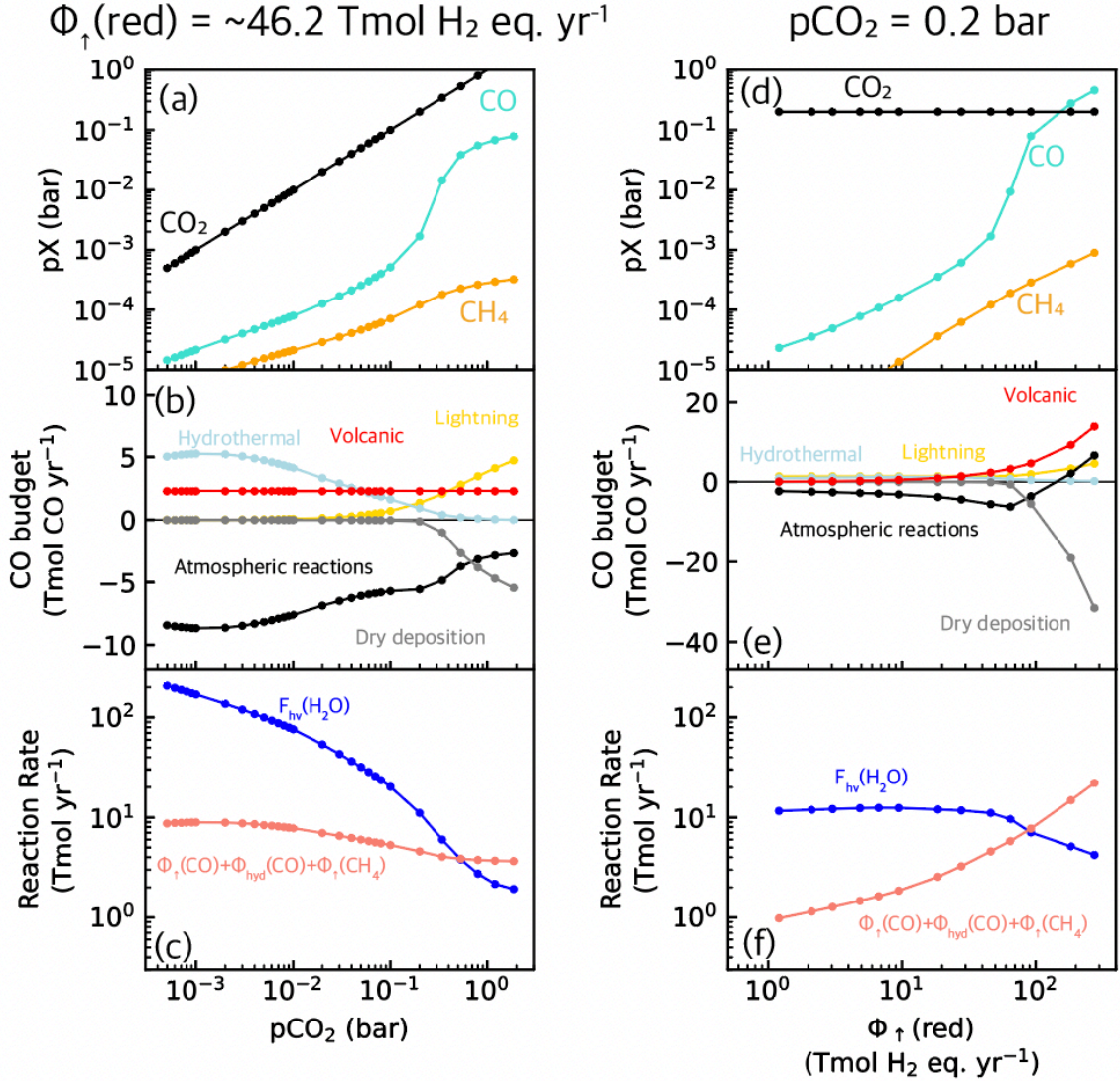

**Figure 3.** Steady-state response of atmospheric chemistry to changes in (left) atmospheric $p$CO$_2$ and (right) $\Phi_\uparrow$(red). The solar spectrum of the young Sun (4 Ga) is assumed. (a and d) Atmospheric composition (blue = CO, black = CO$_2$, cyan = H$_2$, and orange = CH$_4$). (b and e) CO budget in the atmosphere (positive/negative values represent source/sink fluxes). (c and f) Photodissociation rate of H$_2$O (blue) and the input rate of reduced C species (CO and CH$_4$).

## 3.2. Conditions required for CO runaway

We analyzed the above results to better understand the processes that lead to CO runaway. We found that the inception of CO runaway is controlled primarily by the relative flux between OH production from $H_2O$ photodissociation ($F_{hv}(H_2O)$) and the input of reducing carbon species ($\Phi_{\uparrow}(CO) + \Phi_{hyd}(CO) + \Phi_{\uparrow}(CH_4)$) (Figure 3c and 3f). The conditions under which these fluxes balance correspond well to the occurrence of CO runaway. We found that CO runaway occurs when these fluxes become comparable even when the model is run without either CO or $CH_4$ in volcanic gases (Figure B1). In such cases, the values of atmospheric $pCO_2$ and $\Phi_{\uparrow}(red)$ required for triggering CO runaway tend to be higher than those shown in Figure 3. Specifically, assuming no external input of reducing carbon species, CO runaway is not observed at the parameter ranges that we investigated.

The findings of further systematic sensitivity experiments also support the above mechanistic understanding of CO runaway (Figure 4a). The conditions for $F_{hv}(H_2O) = \Phi_{\uparrow}(CO) + \Phi_{hyd}(CO) + \Phi_{\uparrow}(CH_4)$ correspond well to the marked increase in atmospheric $pCO$. The present result also demonstrates that the threshold value of $pCO_2$ for CO runaway decreases with increasing $\Phi_{\uparrow}(red)$, and that elevated $pCO$ levels are accompanied by elevated $pCH_4$ (Figure 4b). Under CO runaway conditions, atmospheric $pCH_4$ levels tend to be high ($\sim > 10^{-4}$ bar) because of the high outgassing rate and reduction in the mixing ratios of OH radicals (Figure C1). Nevertheless, unlike the runaway behavior in atmospheric $pCO$, atmospheric $pCH_4$ does not exhibit such dramatic change.

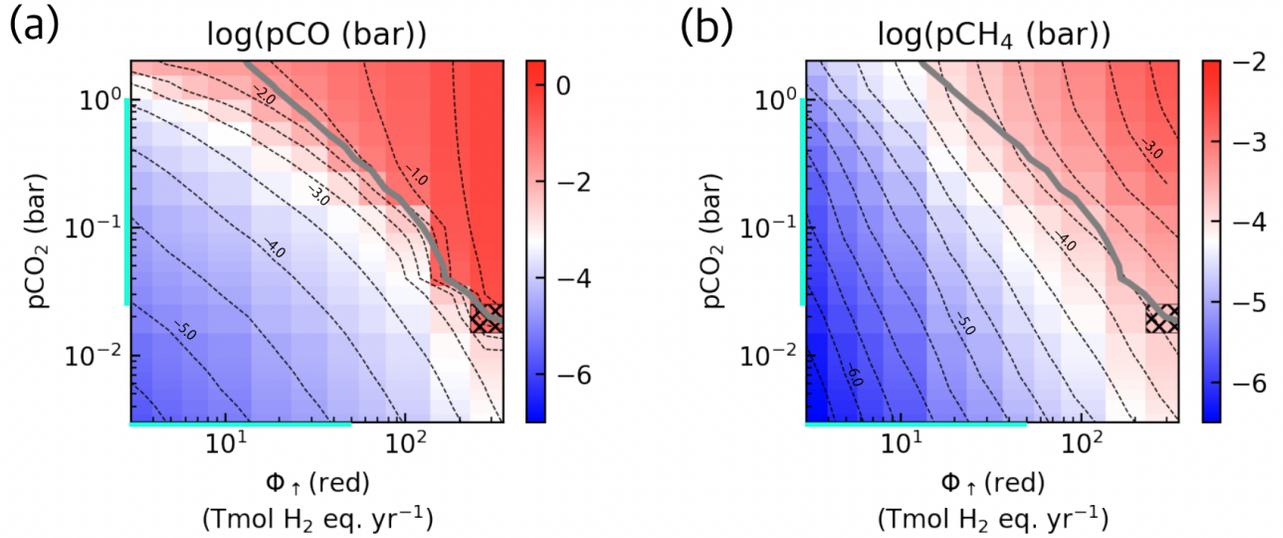

**Figure 4.** Steady-state response of atmospheric CO and CH$_4$ abundances in a phase space of atmospheric $p$CO$_2$ and $\Phi_\uparrow$(red), assuming the solar spectrum of the young Sun (4 Ga). (a) atmospheric $p$CO and (b) atmospheric $p$CH$_4$. Gray line represents the conditions for $F_{hv}$(H$_2$O) = $\Phi_\uparrow$(CO) + $\Phi_{hyd}$(CO) + $\Phi_\uparrow$(CH$_4$), representing the onset of CO runaway. Light-blue bars on $x$ and $y$ axes represent the range of $\Phi_\uparrow$(red) (Krissansen-Totton et al. 2018b) and the range of $p$CO$_2$ for the 4 Ga Earth's atmosphere estimated using a global C cycle model (Krissansen-Totton et al. 2018a), respectively. Area with thick black hatching represents the condition without sufficient convergence.

3.3. Production of organic compounds

In the prebiotic atmosphere, a series of organic compounds are produced in the atmosphere. The largest removal pathway of organic compounds from the atmosphere is deposition of $H_2CO$, which is shown as a function of atmospheric $pCO_2$ and $\Phi_{\uparrow}$(red) (Figure 5). Under CO runaway conditions, the $H_2CO$ deposition rate is smaller than under conditions without CO runaway. When CO runaway occurs, the production rate of H radicals in the troposphere decreases dramatically owing to depletion of OH radicals (see Appendix C), suppressing the production of tropospheric HCO:

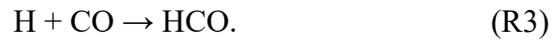

$$H + CO \rightarrow HCO. \qquad\qquad (R3)$$

Consequently, the concentration of tropospheric HCO is decreased under CO runaway conditions, which further leads to decline in the reactions between pairs of HCO molecules that form $H_2CO$:

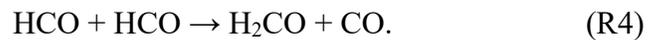

$$HCO + HCO \rightarrow H_2CO + CO. \qquad\qquad (R4)$$

Thus, the rate of formation of tropospheric $H_2CO$ decreases when CO runaway occurs (Figure 5b and 5c). It also leads to decline in the rate of supply of prebiotic organic compounds to the ocean. Under CO runaway conditions, the depositional flux of $H_2CO$ is <1 Tmol C $yr^{-1}$, while that of HCO is typically ~0.01 Tmol C $yr^{-1}$ (Figure 5b and 5c); therefore, it does not strongly affect the atmospheric CO budget (Figure 3b). Instead of these organic compounds, the removal of CO is the primary sink of reduced C in the atmosphere under CO runaway conditions (Figures 3b, 3e, and 5a). This implies that CO chemistry in the aqueous phase is critical for the formation of organic compounds in prebiotic oceans.

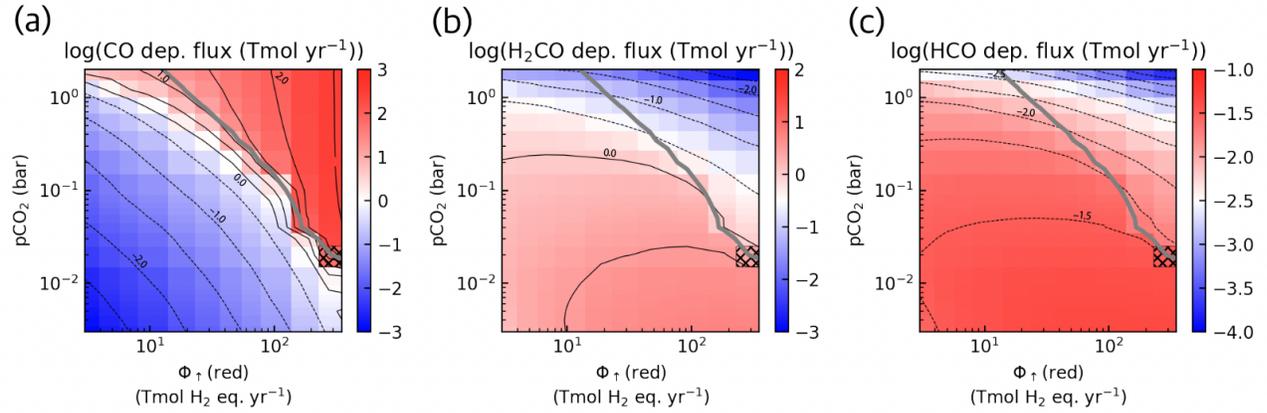

**Figure 5.** Same as Figure 4 but showing the depositional flux of (a) CO, (b) H$_2$CO, and (c) HCO as a function of atmospheric $p$CO$_2$ and $\Phi_\uparrow$(red).

3.4. Dependency on temperature

In this study, we employ a photochemistry model prescribing the temperature and $H_2O$ profiles in the atmosphere. However, the climatic state would affect the conditions for CO runaway because it exerts fundamental control on the availability of OH in the atmosphere. For example, climate warming would lead to increase in $H_2O$ in the troposphere, suppressing CO runaway by promoting production of OH. To examine the impact of uncertainty in temperature (climate) on our results, we conducted additional sensitivity experiments with respect to an elevated surface temperature (300 K at the surface) (Figure 6). As expected, the atmospheric $CO_2$ levels and the outgassing rate of reducing gases ($\Phi_1(red)$) required for CO runaway tend to be larger than in the standard case (cf. Figure 3). Quantifying the interrelationship between photochemistry and climate using a fully coupled model of photochemistry and climate (Arney et al. 2016, 2017, 2018; Garduno Ruiz et al. 2023) will be a fruitful topic for future research.

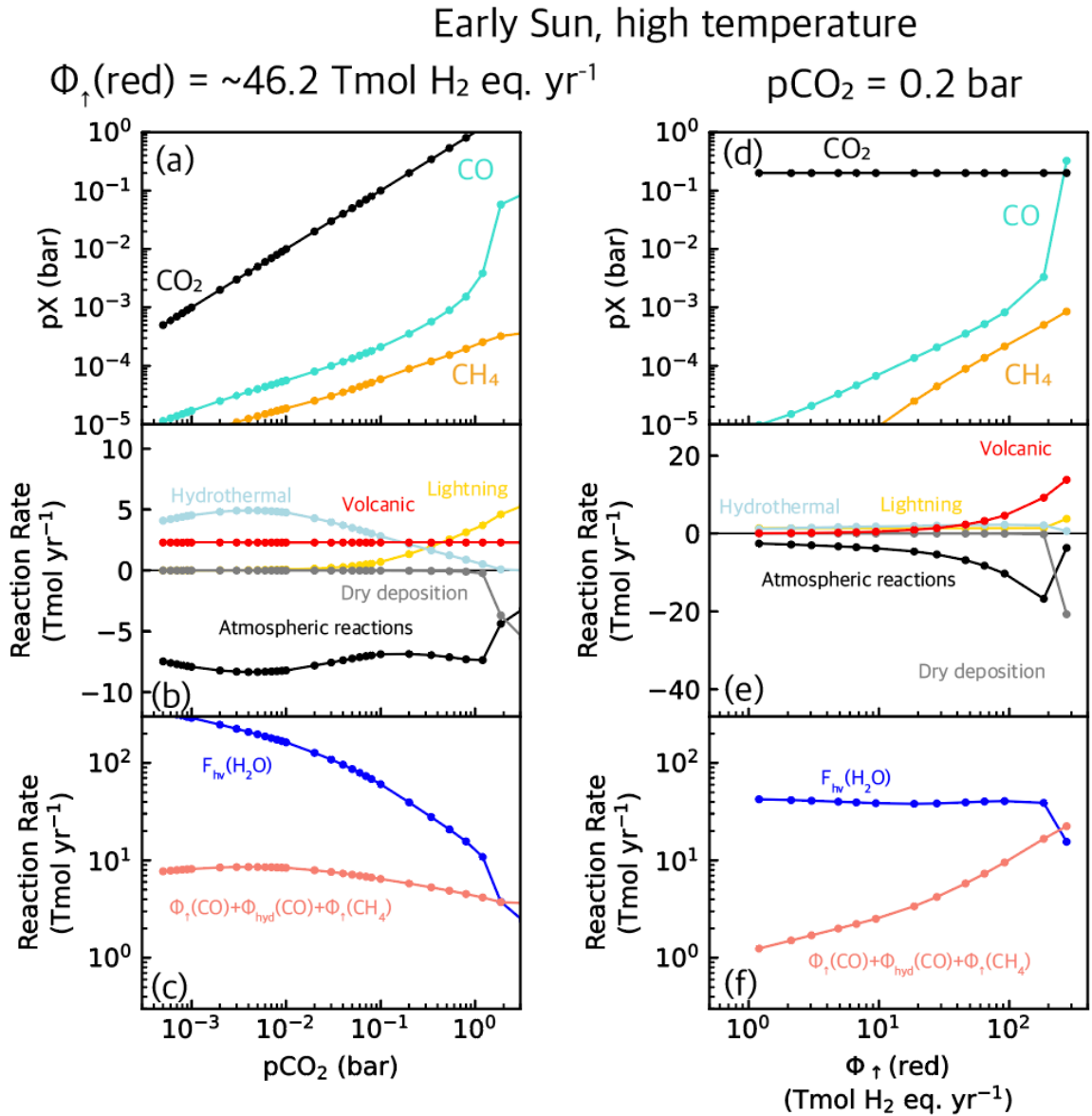

**Figure 6.** Same as Figure 3 but when a higher tropospheric temperature (red line in Figure A1) is assumed.

3.5. Dependency on spectrum of the central stars

The effect of the stellar spectrum type on the response of the atmospheric carbon composition and photodissociation rates of $CO_2$ and $H_2O$ was examined by performing additional experiments with the spectrum of F2V and K2V stars. For the case of the F2V star (Figure 7), CO runaway does not occur even at the maximum value of atmospheric $pCO_2$ that we explored. This can be attributed to the stronger UV flux of the F2V star than that of the Sun. Consequently, the photodissociation rates of $CO_2$ and $H_2O$ are more than one order of magnitude larger than those shown in Figure 3c. This allows elevation of the rate of production of OH in the atmosphere, which suppresses the onset of CO runaway. In contrast, for the case of the K2V star (Figure 8), CO runaway occurs at a lower level of atmospheric $pCO_2$ ($\sim 1$–$2 \times 10^{-3}$ bar) than that for the case of the early Sun. This can be attributed to the relatively small UV flux from the K2V star. Specifically, the UV flux at $\sim 160$–200 nm is smaller than that of the Sun. This suppresses the photodissociation rate of $H_2O$, which promotes the occurrence of CO runaway. These results indicate that the characteristics of the stellar spectra are an important factor controlling the conditions for CO runaway.

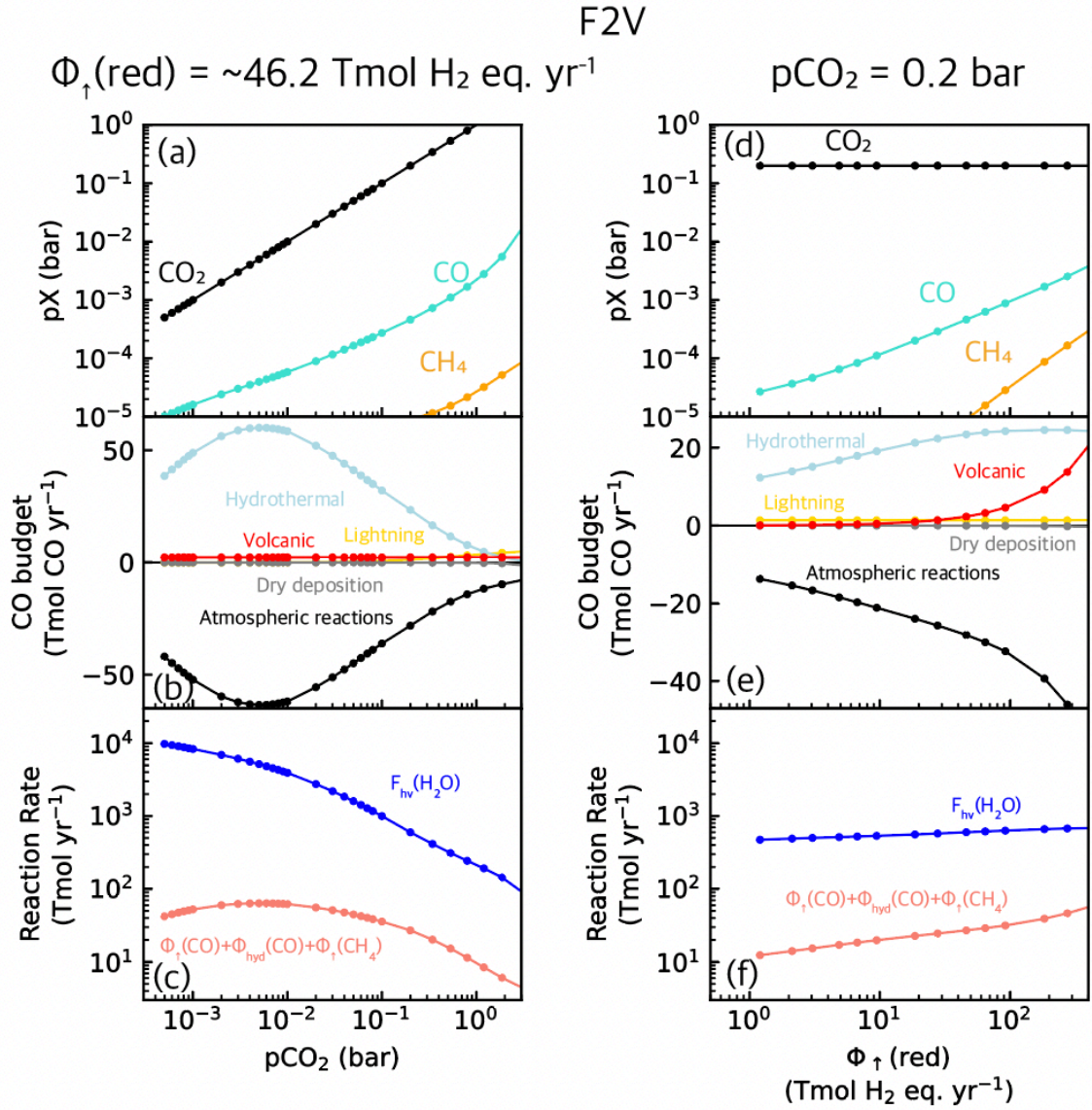

**Figure 7.** Same as Figure 3 but when the stellar spectrum of the F2V star is adopted.

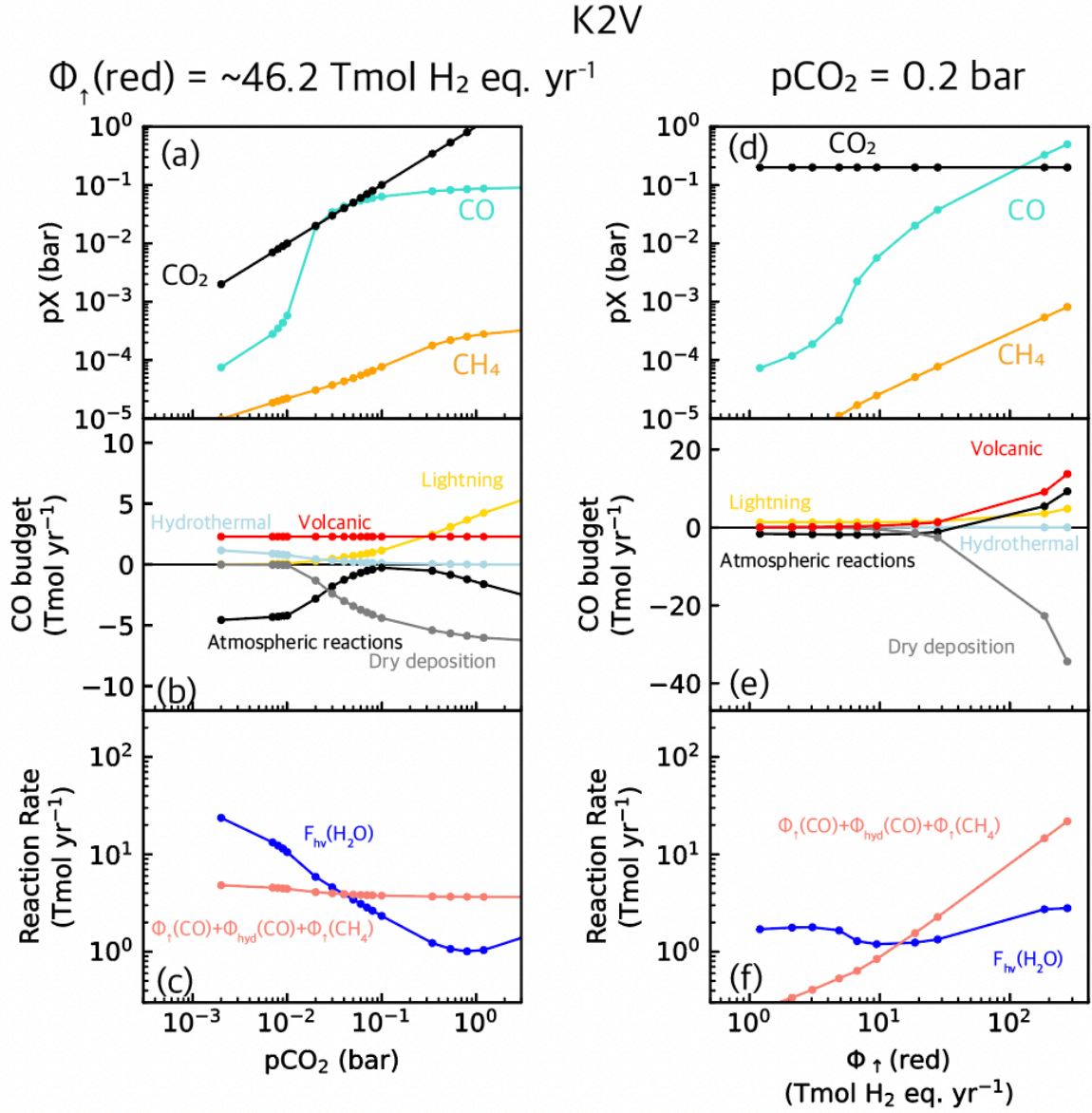

**Figure 8.** Same as Figure 3 but when the stellar spectrum of the K2V star is adopted.

3.6. CO runaway gap

When the simulated atmospheric $pCO_2$, $pCH_4$, and $pCO$ are plotted in the phase space of $pCO/pCO_2$ versus $pCH_4/pCO_2$, the atmospheric chemistry is clearly divided into the groups before and after the occurrence of CO runaway (Figure 9). These relationships, in principle, can be explained by the chemical reactions that determine the atmospheric levels of $pCO$ and $pCH_4$. Before CO runaway, atmospheric $pCO$ and $pCH_4$ are balanced by reactions with OH radicals. Under such conditions, the mixing ratios of CO and CH4 ($f_{CO}$ and $f_{CH4}$, respectively) can be approximated as follows:

$$f_{CO} \simeq \frac{\Phi_{\uparrow}(CO) + \Phi_{hyd}(CO)}{\int K_{CO+OH} \cdot n_{OH}(z) \cdot n_{air}(z) \cdot dz} \quad , \qquad (8)$$

$$f_{CH4} \simeq \frac{\Phi_{\uparrow}(CH_4)}{\int K_{CH4+OH} \cdot n_{OH}(z) \cdot n_{air}(z) \cdot dz}, \qquad (9)$$

where $K_{CO+OH}$ and $K_{CH4+OH}$ are the reaction rate coefficients for reaction R11 and R12 in Appendix C, respectively, and $n_{OH}$ and $n_{atm}$ represent the number density of OH radicals and the air, respectively. By neglecting the temperature- and pressure-dependencies of $K_{CO+OH}$ and $K_{CH4+OH}$, the CH4/CO2 and CO/CO2 relationship can be approximated as follows:

$$\frac{CH_4/CO_2}{CO/CO_2} = \frac{pCH_4}{pCO} = \frac{f_{CH4}}{f_{CO}} \simeq \frac{K_{CH4+OH} \cdot \Phi_{\uparrow}(CH_4)}{K_{CO+OH} \cdot (\Phi_{\uparrow}(CO) + \Phi_{hyd}(CO))}. \qquad (10)$$

This equation indicates that CH4/CO2 and CO/CO2 would have almost linear dependence because the outgassing rates of CO and abiotic CH4 would also be linearly dependent.

Under CO runaway conditions, the reactions that consume atmospheric CO are different from those before CO runaway. In such cases, the CO mixing ratio in the atmosphere could be represented approximately as follows:

$$f_{CO} \simeq \frac{\Phi_{\uparrow}(CO) + \Phi_{hyd}(CO)}{v_{dep}(CO) \cdot n_{atm}(z=0) + \int K_{CO+O} \cdot n_O(z) \cdot n_{atm}(z) \cdot dz} \quad , \qquad (11)$$

where $K_{CO+O}$ is the reaction rate coefficient for reaction R13 (Appendix C). The high $f_{CO}$ is caused by the high CO outgassing rate and/or low value in the denominator. When CO runaway occurs in the atmosphere, the atmospheric $CH_4$ cycle tends to be balanced by photodissociation at high altitudes ($\sim>80$ km) because of insufficient OH radicals at lower altitudes. Because photodissociation of $CH_4$ effectively consumes $CH_4$, the sudden increase in atmospheric $pCH_4$ is not caused by the scarcity of OH radicals when CO runaway occurs in the atmosphere. This would suppress the change in atmospheric $pCH_4$ compared with the atmospheric $pCO$, leading to the gap structure shown in Figure 9. When the outgassing rate of $CH_4$ is small, the CO runaway branch does not exhibit a clear structure. Nevertheless, the sudden increase in atmospheric $pCO$ contributes to the gap structure (Figure B2).

This gap structure originates from the different behavior of atmospheric CO and $CH_4$, and therefore it might be a ubiquitous feature for various Earth-like exoplanets (Figure 9). Indeed, compared with the case of the G2V star, the gap structure is clearly seen in the case of the K2V star because of the ease with which CO runaway is caused. For the parameter ranges that we explored, most calculations for the case of the F2V star lay on the left branch because it is difficult to cause CO runaway with a high $H_2O$ photodissociation rate (Figure 7c and 7f). Nonetheless, some results did demonstrate CO runaway, especially for high atmospheric $pCO_2$ and/or elevated $\Phi_\uparrow$(red) cases, representing the existence of the CO runaway gap structure.

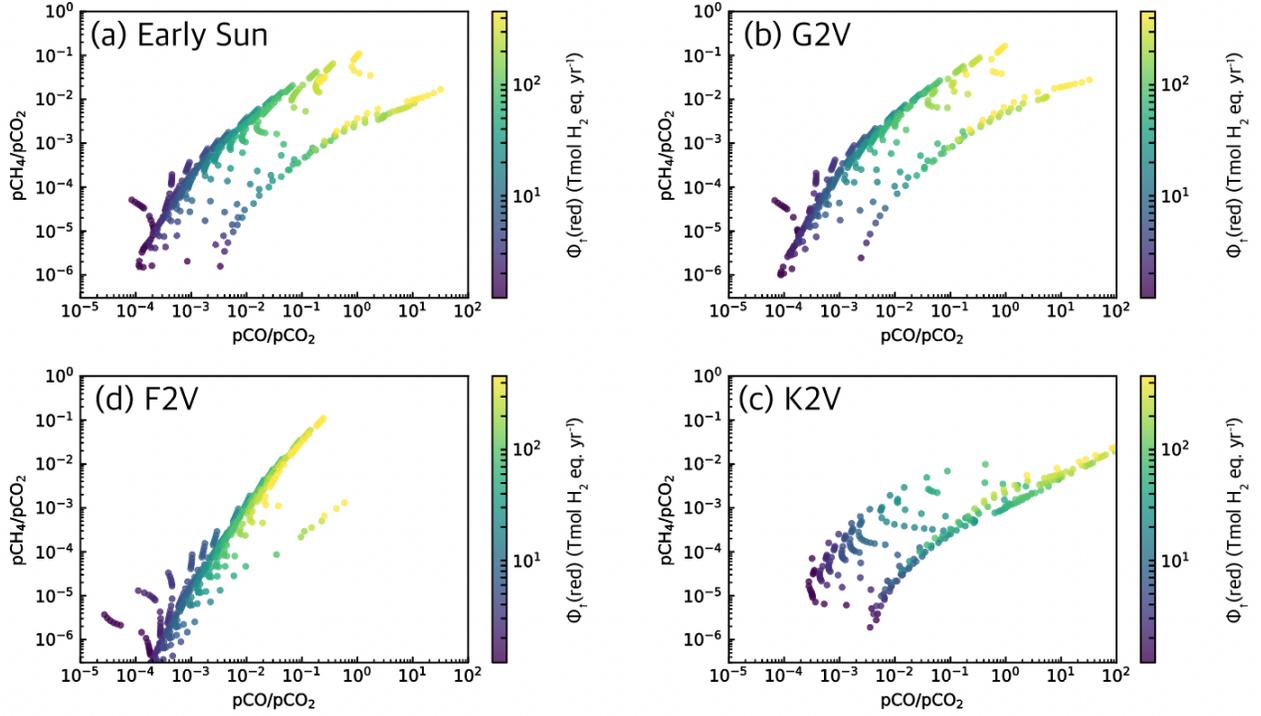

**Figure 9.** Steady-state response of atmospheric $CH_4/CO_2$ and $CO/CO_2$ for different types of central stars: (a) the young Sun (4 Ga), (b) the present Sun (G2V), (c) the σ Bootis (F2V), and (d) the ε Eridani (K2V). Colors represent the volcanic outgassing flux of reducing gases, $\Phi_{\uparrow}$(red).

## 4. DISCUSSION

4.1. Possibility of CO runaway on the early Earth and Mars

The atmospheric $p\text{CO}_2$ level and the volcanic input flux of reducing carbon species (CO and $\text{CH}_4$) exert fundamental controls on the conditions for CO runaway by changing the photodissociation rate of water in the atmosphere (Figure 4). For our standard case, the critical $p\text{CO}_2$ level for CO runaway is a few tenths of one bar for the outgassing rate of reducing gases of a few tens Tmol $\text{H}_2$ equivalent $\text{yr}^{-1}$. In terms of atmospheric $\text{CO}_2$ levels, this would be feasible for the early Earth because the steady-state atmospheric $p\text{CO}_2$ would have been elevated via the carbonate–silicate geochemical cycle owing to the low solar luminosity (Gough 1981; Walker et al. 1981; Tajika and Matsui 1992; Sleep and Zahnle 2001; Krissansen-Totton et al. 2018a; Kadoya and Tajika 2019; Kadoya et al. 2020; Lehmer et al. 2020; Watanabe and Tajika 2021). Specifically, previous theoretical models of the global carbon cycle have estimated steady-state atmospheric $p\text{CO}_2$ levels of ~0.1–1 bar or higher during the Hadean–early Archean (Tajika and Matsui 1992; Sleep and Zahnle 2001; Krissansen-Totton et al. 2018a).

An additional condition for CO runaway is a high volcanic outgassing rate of reducing gases, $\Phi_\uparrow(\text{red})$. Previous studies estimated that $\Phi_\uparrow(\text{red})$ would typically be ~1 Tmol $\text{H}_2$ eq. $\text{yr}^{-1}$ or less on the early Earth (Holland 1984; Catling and Kasting 2017; Thompson et al. 2022), while the maximum outgassing flux from serpentinization is estimated to be ~50 Tmol $\text{H}_2$ eq. $\text{yr}^{-1}$ (Krissansen-Totton et al. 2018b). Our results demonstrate that a value of $\Phi_\uparrow(\text{red})$ of ~>10 Tmol $\text{H}_2$ eq. $\text{yr}^{-1}$ is required for CO runaway for a level of atmospheric $p\text{CO}_2$ of ~<1 bar (Figure 4). This is within the possible maximum value estimated by Krissansen-Totton et al. (2018). From a more mechanistic perspective, the relative outgassing fluxes of reduced C (CO and $\text{CH}_4$) would be critical for the occurrence of CO runaway (Appendix B). The outgassing rates of CO and $\text{CH}_4$ relative to $H_2$ reflect the conditions of the thermodynamic equilibrium between magma and gas bubbles. Specifically, the high-pressure conditions for submarine volcanism would be preferable for the outgassing of $CH_4$ relative to $H_2$ (Gaillard et al. 2011; Wogan et al. 2020), promoting CO runaway. The high-temperature conditions and high oxygen fugacity in magma would have promoted the outgassing of CO relative to $CH_4$ (Wogan et al. 2020). Therefore, for better constraint on the conditions required for CO runaway, coupled modeling of atmospheric and solid-Earth processes is critical. In addition to the outgassing flux from the interior of the Earth, impact events would have played a crucial role in providing CO on early Earth, at least transiently (Kasting 1990, 2014; Kress and McKay 2004; Zahnle 2006; Schaefer and Fegley 2007; Genda et al. 2017; Zahnle et al. 2020). This would also have helped initiate CO runaway in the early atmosphere. Nevertheless, our results clearly demonstrate the conditions for CO runaway in terms of the levels of atmospheric $CO_2$ and $\Phi_\uparrow(\mathrm{red})$ (Figure 4).

Other processes not yet fully included in the model warrant future consideration, e.g., the effect of climate on the conditions for CO runaway. Variation in the planetary climate and the associated change in the water mass in the troposphere would play a role in the conditions for CO runaway by changing the availability of OH in the troposphere (Figure 6). The uncertainty in the climatic state does not alter our overarching conclusions regarding the fundamental role of atmospheric $CO_2$ and $\Phi_\uparrow(\mathrm{red})$, and the gap structure in the $p\mathrm{CH_4}/p\mathrm{CO_2}$–$p\mathrm{CO}/p\mathrm{CO_2}$ phase space. However, complete treatment of the coupled relationship between atmospheric chemistry and the climate will be essential for quantitative understanding of how the relative abundances of $CO_2$, CO, and $CH_4$ affect planetary habitability and (bio)geochemistry, thereby representing a fruitful topic for future research.

The effect of climate on the relative abundances of $CO_2$, CO, and $CH_4$ in the atmosphere would be even more critical for the case of Mars. The climate of early Mars has been a subject of active discussion (Pollack et al. 1987; Kasting 1991; Forget et al. 2013; Wordsworth 2016; Wordsworth et al. 2017; Ramirez and Craddock 2018; Kamada et al. 2020). If a cold and dry environment were pervasive on early Mars, it might have promoted CO runaway if the level of atmospheric $p$CO$_2$ was sufficiently high. In contrast, if a warm and wet environment were pervasive, a higher atmospheric $p$CO$_2$ level would have been required. The recent findings of anomalously low negative carbon isotopic values of the Martian atmospheric CO and organic matter buried in Gale crater suggest the importance of CO production via $CO_2$ photolysis in the modern and past Martian atmosphere (Zahnle et al. 2008; Krasnopolsky 2015; House et al. 2022; Yoshida et al. 2023; Alday et al. 2023; Aoki et al. 2023), opening new vistas for future work intended to better understand the possibility of CO runaway on early Mars.

## 4.2. Implications for the origin of life

Studies have discussed that CO runaway conditions promote the formation of organic compounds, aldehydes, peptides, and/or amino acids suitable for prebiotic chemistry (Bar-Nun and Chang 1983; Huber and Wächtershäuser 1997, 1998; McGlynn et al. 2020; Zang et al. 2022). Our model demonstrates that elevated atmospheric $CO_2$ levels and/or the high volcanic outgassing flux of reducing gases promote the formation of CO-rich atmospheres suitable for the origin of life. Our results also indicate that the rate of CO deposition is exceptionally high when CO runaway occurs (Figures 4b, 4d, and 5a). This would have also played a key role in supplying reduced C species to the ocean, which might indicate that the oceanic chemistry of CO and other organic compounds is crucial in understanding prebiotic chemistry. The CO concentrations in the ocean, [CO], can be estimated using the simple oceanic chemistry model constructed by Kharecha et al. (2005) (Appendix D). The results demonstrate that under CO runaway conditions, [CO] reaches an order of 100 $\mu$M (blue and light blue lines in Figure D1a and D1b). We can also estimate the maximum concentration of $H_2CO$ in the ocean (Figure D1c). Although the deposition flux of $H_2CO$ decreases under CO runaway conditions (Figure 5b), the maximum concentration of $H_2CO$ in the ocean would exceed 1 mM, which would have helped the formation of the building blocks of life together with CO in the ocean.

## 4.3. Implications for the search for habitable planets

Our results demonstrate that anoxic atmospheres of Earth-like lifeless planets can be clearly classified in the $p$CH$_4$/$p$CO$_2$–$p$CO/$p$CO$_2$ phase space. The sensitivity experiments with respect to different types of central stars also indicate that the CO runaway gap structure would be a general feature of Earth-like planets orbiting Sun-like stars. Our results indicate that the most preferable condition for CO runaway is near the outer edge of the HZ because atmospheric $p$CO$_2$ is expected to be high in this region owing to the operation of carbonate–silicate geochemical cycles (Walker et al. 1981; Kadoya and Tajika 2019; Kadoya et al. 2020; Lehmer et al. 2020; Watanabe and Tajika 2021). In this study, we did not consider planets orbiting M-type stars. Nevertheless, because such planets receive less near-UV radiation than planets orbiting Sun-like stars, we expect that CO runaway would tend to be promoted in the atmospheres of such planets. More systematic examinations that comprehensively consider astronomical factors and planetary endogenous factors will be the subject of future work.

## 5. CONCLUSION

Understanding the factors that control the relative abundances of $CO_2$, CO, and $CH_4$ in planetary atmospheres has important implications in the search for habitable planets beyond our solar system because atmospheric composition exerts fundamental controls on planetary climate and biogeochemistry. In this study, we employed an atmospheric photochemistry model to understand the diversity of atmospheric $CO_2$, CO, and $CH_4$ abundances in the atmosphere of Earth-like planets orbiting Sun-like stars. Our results for the prebiotic Earth conditions demonstrate that photochemical instability of CO (i.e., CO runaway) tends to be triggered by higher atmospheric $p$$CO_2$ levels and increased outgassing fluxes of reducing gases. Our systematic examinations reveal that atmospheric chemistry can be clearly classified in the phase space of $p$$CH_4$/$p$$CO_2$ versus $p$CO/$p$$CO_2$, where a distinct gap structure appears. Our results also demonstrate that this CO runaway gap structure would be a general feature of Earth-like planets orbiting Sun-like stars, providing insight into the characteristics and potential habitability of exoplanets. These results advance our understanding of the intricate relationship between the spectral types of central stars, atmospheric composition, climate, tectonic activity, and the origin of life, contributing to the ongoing exploration of habitable worlds beyond our solar system.

## 6. Acknowledgments

We thank E. W. Schwieterman for his assistance in adapting molecular cross-sections from the *Atmos* model. We also thank C. T. Reinhard, Y. Ueno, A. Akahori, and E. Tajika for their helpful discussions. This work was supported by JSPS KAKENHI grant numbers 22H05149 and 22H05150, JST FOREST Program (Grant Number JPMJFR23XX, Japan), and Mitsubishi Foundation (202210014). K.O. acknowledges the NASA Nexus for Exoplanet System Science (NExSS). We thank James Buxton MSc, from Edanz (https://jp.edanz.com/ac), for editing a draft of this manuscript.

## 7. Author Contribution

Conceptualization: KO, YW; Methodology: YW, KO; Software: YW, KO; Supervision: KO; Investigation: KO, YW; Visualization: YW; Writing – original draft: YW, KO; Writing – review & editing: KO, YW; Funding acquisition: KO

## 8. Competing Interests

The authors declare no competing interest.

## References

Abe Y, Abe-Ouchi A, Sleep NH, Zahnle KJ (2011) Habitable zone limits for dry planets. Astrobiology 11:443–460

Alday J, Trokhimovskiy A, Patel MR, et al (2023) Photochemical depletion of heavy CO isotopes in the Martian atmosphere. Nature Astronomy 1–10

Anglada-Escudé G, Amado PJ, Barnes J, et al (2016) A terrestrial planet candidate in a temperate orbit around Proxima Centauri. Nature 536:437–440

Aoki S, Shiobara K, Yoshida N, et al (2023) Depletion of 13C in CO in the Atmosphere of Mars Suggested by ExoMars-TGO/NOMAD Observations. Planet Sci J 4:97

Arney G, Domagal-Goldman SD, Meadows VS, et al (2016) The Pale Orange Dot: The Spectrum and Habitability of Hazy Archean Earth. Astrobiology 16:873–899

Arney G, Domagal-Goldman SD, Meadows VS (2018) Organic Haze as a Biosignature in Anoxic Earth-like Atmospheres. Astrobiology 18:311–329

Arney GN, Meadows VS, Domagal-Goldman SD, et al (2017) Pale Orange Dots: The Impact of Organic Haze on the Habitability and Detectability of Earthlike Exoplanets. Astrophys J 836:1–19

Bar-Nun A, Chang S (1983) Photochemical reactions of water and carbon monoxide in Earth's primitive atmosphere. J Geophys Res 88:6662

Bartik K, Bruylants G, Locci E, et al (2011) Liquid water: a necessary condition for all forms of life. Origins and Evolution of life: an Astrobiological Perspective 205–217

Catling DC, Kasting JF (2017) Atmospheric evolution on inhabited and lifeless worlds. Cambridge University Press

Cleaves HJ (2008) The prebiotic geochemistry of formaldehyde. Precambrian Res 164:111–118

Forget F, Wordsworth R, Millour E, et al (2013) 3D modelling of the early martian climate under a denser $CO_2$ atmosphere: Temperatures and $CO_2$ ice clouds. Icarus 222:81–99

Gaillard F, Scaillet B, Arndt NT (2011) Atmospheric oxygenation caused by a change in volcanic degassing pressure. Nature 478:229–232

Garduno Ruiz D, Goldblatt C, Ahm A-S (2023) Climate shapes the oxygenation of Earth's atmosphere across the Great Oxidation Event. Earth Planet Sci Lett 607:118071

Genda H, Brasser R, Mojzsis SJ (2017) The terrestrial late veneer from core disruption of a lunar-sized impactor. Earth Planet Sci Lett 480:25–32

Gillon M, Triaud AHMJ, Demory B-O, et al (2017) Seven temperate terrestrial planets around the nearby ultracool dwarf star TRAPPIST-1. Nature 542:456–460

Gough DO (1981) Solar Interior Structure and Luminosity Variations. In: Physics of Solar Variations. Springer Netherlands, pp 21–34

Harman CE, Kasting JF, Wolf ET (2013) Atmospheric production of glycolaldehyde under hazy prebiotic conditions. Orig Life Evol Biosph 43:77–98

Hart MH (1979) Habitable zones about main sequence stars. Icarus 37:351–357

Hill ML, Bott K, Dalba PA, et al (2023) A Catalog of Habitable Zone Exoplanets. AJS 165:34

Holland HD (1984) The chemical evolution of the atmosphere and oceans. Princeton University Press

House CH, Wong GM, Webster CR, et al (2022) Depleted carbon isotope compositions observed at Gale crater, Mars. Proc Natl Acad Sci U S A 119.: https://doi.org/10.1073/pnas.2115651119

Huang S-S (1959) OCCURRENCE OF LIFE IN THE UNIVERSE. Am Sci 47:397–402

Huber C, Wächtershäuser G (1997) Activated acetic acid by carbon fixation on (Fe,Ni)S under primordial conditions. Science 276:245–247

Huber C, Wächtershäuser G (1998) Peptides by activation of amino acids with CO on (Ni,Fe)S surfaces: implications for the origin of life. Science 281:670–672

Hu R, Peterson L, Wolf ET (2020) O2- and CO-rich Atmospheres for Potentially Habitable Environments on TRAPPIST-1 Planets. ApJ 888:122

Kadoya S, Krissansen-Totton J, Catling DC (2020) Probable cold and alkaline surface environment of the hadean earth caused by impact ejecta weathering. Geochem Geophys Geosyst 21.: https://doi.org/10.1029/2019gc008734

Kadoya S, Tajika E (2019) Outer Limits of the Habitable Zones in Terms of Climate Mode and Climate Evolution of Earth-like Planets. ApJ 875:7

Kaltenegger L, Sasselov D (2011) EXPLORING THE HABITABLE ZONE FOR KEPLER PLANETARY CANDIDATES. ApJL 736:L25

Kamada A, Kuroda T, Kasaba Y, et al (2020) A coupled atmosphere–hydrosphere global climate model of early Mars: A "cool and wet" scenario for the formation of water channels. Icarus 338:113567

Kasting JF (2014) Atmospheric composition of Hadean–early Archean Earth: The importance of CO. Geological Society of America Special Papers 504:19–28

Kasting JF (1990) Bolide impacts and the oxidation state of carbon in the Earth's early atmosphere. Origins of Life and Evolution of the Biosphere 20:199–231

Kasting JF (1991) CO2 condensation and the climate of early Mars. Icarus 94:1–13

Kasting JF, Kopparapu R, Ramirez RM, Harman CE (2014) Remote life-detection criteria, habitable zone boundaries, and the frequency of Earth-like planets around M and late K stars. Proc Natl Acad Sci U S A 111:12641–12646

Kasting JF, Whitmire DP, Reynolds RT (1993) Habitable zones around main sequence stars. Icarus 101:108–128

Kasting JF, Zahnle KJ, Walker JCG (1983) Photochemistry of methane in the Earth's early atmosphere. Precambrian Res 20:121–148

Kharecha P, Kasting J, Siefert J (2005) A coupled atmosphere-ecosystem model of the early

Archean Earth. Geobiology 3:53–76

Kopparapu RK, Ramirez R, Kasting JF, et al (2013) HABITABLE ZONES AROUND MAIN-SEQUENCE STARS: NEW ESTIMATES. The Astrophysical Journal 2013.: https://doi.org/10.1088/0004-637X/765/2/131

Kopparapu RK, Ramirez RM, SchottelKotte J, et al (2014) HABITABLE ZONES AROUND MAIN-SEQUENCE STARS: DEPENDENCE ON PLANETARY MASS. ApJL 787:L29

Krasnopolsky VA (2015) Variations of carbon monoxide in the martian lower atmosphere. Icarus 253:149–155

Kress ME, McKay CP (2004) Formation of methane in comet impacts: implications for Earth, Mars, and Titan. Icarus 168:475–483

Krissansen-Totton J, Arney GN, Catling DC (2018a) Constraining the climate and ocean pH of the early Earth with a geological carbon cycle model. Proc Natl Acad Sci U S A 115:4105–4110

Krissansen-Totton J, Olson S, Catling DC (2018b) Disequilibrium biosignatures over Earth history and implications for detecting exoplanet life. Sci Adv 4:eaao5747

Lehmer OR, Catling DC, Krissansen-Totton J (2020) Carbonate-silicate cycle predictions of Earth-like planetary climates and testing the habitable zone concept. Nat Commun 11:6153

Lincowski AP, Meadows VS, Crisp D, et al (2018) Evolved Climates and Observational Discriminants for the TRAPPIST-1 Planetary System. ApJ 867:76

Mayor M, Queloz D (1995) A Jupiter-mass companion to a solar-type star. Nature 378:355–359

McGlynn SE, Glass JB, Johnson-Finn K, et al (2020) Hydrogenation reactions of carbon on Earth: Linking methane, margarine, and life. Am Mineral 105:599–608

Meadows VS, Reinhard CT, Arney GN, et al (2018) Exoplanet Biosignatures: Understanding Oxygen as a Biosignature in the Context of Its Environment. Astrobiology 18:630–662

Miller SL, Urey HC (1959) Organic Compound Synthesis on the Primitive Earth. Science 130:245–251

Pavlov AA, Brown LL, Kasting JF (2001) UV shielding of NH3 and O2 by organic hazes in the Archean atmosphere. J Geophys Res 106:23267–23287

Pollack JB, Kasting JF, Richardson SM, Poliakoff K (1987) The case for a wet, warm climate on early Mars. Icarus 71:203–224

Ragsdale SW (2004) Life with carbon monoxide. Crit Rev Biochem Mol Biol 39:165–195

Ramirez RM, Craddock RA (2018) The geological and climatological case for a warmer and wetter early Mars. Nat Geosci 11:230–237

Ranjan S, Schwieterman EW, Harman C, et al (2020) Photochemistry of Anoxic Abiotic Habitable Planet Atmospheres: Impact of New H2O Cross Sections. The Astrophysical Journal 896:148

Ranjan S, Schwieterman EW, Leung M, et al (2023) A Re-Appraisal of $CO/O_2$ Runaway on Habitable Planets Orbiting Low-Mass Stars. arXiv [astro-ph.EP]

Ranjan S, Seager S, Zhan Z, et al (2022) Photochemical Runaway in Exoplanet Atmospheres: Implications for Biosignatures. ApJ 930:131

Sagan C, Chyba C (1997) The early faint sun paradox: organic shielding of ultraviolet-labile greenhouse gases. Science 276:1217–1221

Schaefer L, Fegley B (2007) Outgassing of ordinary chondritic material and some of its implications for the chemistry of asteroids, planets, and satellites. Icarus 186:462–483

Schmidt JA, Johnson MS, Schinke R (2013) Carbon dioxide photolysis from 150 to 210 nm: singlet and triplet channel dynamics, UV-spectrum, and isotope effects. Proc Natl Acad Sci U S A 110:17691–17696

Schwieterman EW, Reinhard CT, Olson SL, et al (2019) Rethinking CO Antibiosignatures in the Search for Life Beyond the Solar System. ApJ 874:9

Seager S, Schrenk M, Bains W (2012) An astrophysical view of Earth-based metabolic biosignature gases. Astrobiology 12:61–82

Segura A, Krelove K, Kasting JF, et al (2003) Ozone concentrations and ultraviolet fluxes on Earth-like planets around other stars. Astrobiology 3:689–708

Selsis F, Kasting JF, Levrard B, et al (2007) Habitable planets around the star Gliese 581? Astron Astrophys Suppl Ser 476:1373–1387

Sleep NH, Zahnle K (2001) Carbon dioxide cycling and implications for climate on ancient Earth. J Geophys Res 106:1373–1399

Spiegel DS, Menou K, Scharf CA (2008) Habitable Climates. The Astrophysical Journal 681:1609–1623

Tajika E, Matsui T (1992) Evolution of terrestrial proto-CO2 atmosphere coupled with thermal history of the earth. Earth Planet Sci Lett 113:251–266

Thompson MA, Krissansen-Totton J, Wogan N, et al (2022) The case and context for atmospheric methane as an exoplanet biosignature. Proc Natl Acad Sci U S A 119:e2117933119

Thuillier G, Floyd L, Woods TN, et al (2004) Solar irradiance reference spectra for two solar active levels. Adv Space Res 34:256–261

Toon OB, McKay CP, Ackerman TP, Santhanam K (1989) Rapid calculation of radiative heating rates and photodissociation rates in inhomogeneous multiple scattering atmospheres. Journal of Geophysical Research 94:16287

Ueno Y, Schmidt J, Johnson M, et al (2022) Anomalously 13C-depleted organic matter from CO in early Mars atmosphere

Underwood DR, Jones BW, Sleep PN (2003) The evolution of habitable zones during stellar lifetimes and its implications on the search for extraterrestrial life. Int J Astrobiology 2:289–299

Walker JCG, Hays PB, Kasting JF (1981) A negative feedback mechanism for the long-term stabilization of Earth's surface temperature. J Geophys Res 86:9776

Walker JCG, Klein C, Schidlowski M, et al (1983) Environmental evolution of the Archean-Early Proterozoic earth. In: IN: Earth's earliest biosphere: Its origin and evolution (A84-43051 21-51). Princeton. ui.adsabs.harvard.edu, pp 260–290

Watanabe Y, Tajika E (2021) Atmospheric oxygenation of the early earth and earth-like planets driven by competition between land and seafloor weathering. Earth Planets Space 73:1–10

Watanabe Y, Tajika E, Ozaki K (2023) Biogeochemical transformations after the emergence of oxygenic photosynthesis and conditions for the first rise of atmospheric oxygen. Geobiology 21:537–555

Williams DM, Kasting JF (1997) Habitable planets with high obliquities. Icarus 129:254–267

Wogan N, Krissansen-Totton J, Catling DC (2020) Abundant Atmospheric Methane from Volcanism on Terrestrial Planets Is Unlikely and Strengthens the Case for Methane as a Biosignature. The Planetary Science Journal 1:58

Wordsworth RD (2016) The Climate of Early Mars. Annu Rev Earth Planet Sci 44:381–408

Wordsworth R, Kalugina Y, Lokshtanov S, et al (2017) Transient reducing greenhouse warming on early Mars. Geophys Res Lett 44:665–671

Yoshida T, Aoki S, Ueno Y, et al (2023) Strong Depletion of 13C in CO Induced by Photolysis of CO2 in the Martian Atmosphere, Calculated by a Photochemical Model. Planet Sci J 4:53

Zahnle K, Haberle RM, Catling DC, Kasting JF (2008) Photochemical instability of the ancient Martian atmosphere. J Geophys Res 113:E11004

Zahnle KJ (1986) Photochemistry of methane and the formation of hydrocyanic acid (HCN) in the Earth's early atmosphere. Journal of Geophysical Research 91:2819

Zahnle KJ (2006) Earth's Earliest Atmosphere. Elements 2:217–222

Zahnle K, Lupu R, Catling DC, Wogan N (2020) Creation and evolution of impact-generated

reduced atmospheres of early Earth. The Planetary Science 1:11

Zang X, Ueno Y, Kitadai N (2022) Photochemical Synthesis of Ammonia and Amino Acids from Nitrous Oxide. Astrobiology 22:387–398

APPENDIX A

**Temperature profile assumed in the atmospheric photochemical model**

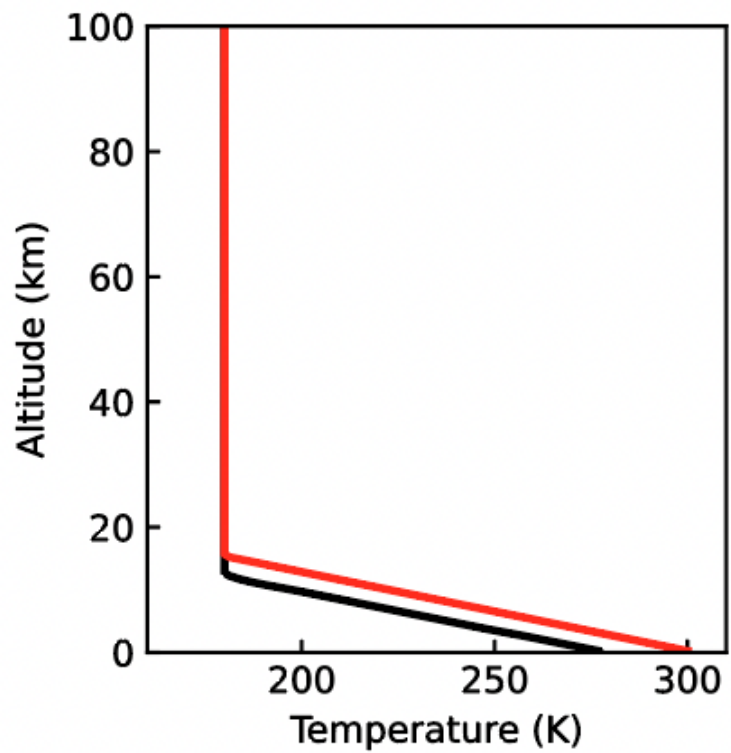

**Figure A1.** The temperature profile adopted in this study: the standard profile (black line; Pavlov et al., 2001; Watanabe et al., 2023) and the profile with high tropospheric temperature (red line).

APPENDIX B

**Effect of the outgassing flux of CO and CH₄**

The occurrence of the CO runaway is controlled by the relative supply rate of OH from the photodissociation of $H_2O$ and reduced C (CO and $CH_4$) from the volcanoes and hydrothermal systems. The difference between the photodissociation rate of $H_2O$ and the external supply rate of reduced C when the model is run without either CO or abiotic $CH_4$ in volcanic gas is shown in Figure B1.

When $CH_4$ enters the atmosphere, it is further converted to $CH_3$, which is further converted to formaldehyde, as follows:

$$CH_3 + O \rightarrow H_2CO + H \qquad (R5)$$

Formaldehyde further reacts with O radicals in the atmosphere to form HCO at high altitudes (~60–80 km):

$$H_2CO + O \rightarrow HCO + OH \qquad (R6)$$

Because this reaction proceeds at high altitudes, it does not affect the tropospheric budget of $H_2CO$ and HCO. The combination of the change in the reaction rates of R4–R6 works as a net conversion of $H_2CO$ to HCO in the atmosphere, leading to the net production of CO via the change in the reaction rate of R3 and following reaction:

$$HCO + O \rightarrow CO + OH \qquad (R7)$$

For this reason, the supply rate of $CH_4$ affects the onset of CO runaway in the same way as the CO is supplied into the atmosphere.

If there is no reduced C outgassing from volcanoes (that is, it is limited to only CO supply from hydrothermal systems), the runaway behavior is not observed because the $H_2O$ photodissociation rate and the reduced C outgassing rate does not get closer (dotted lines in Figure B2a–B2d).

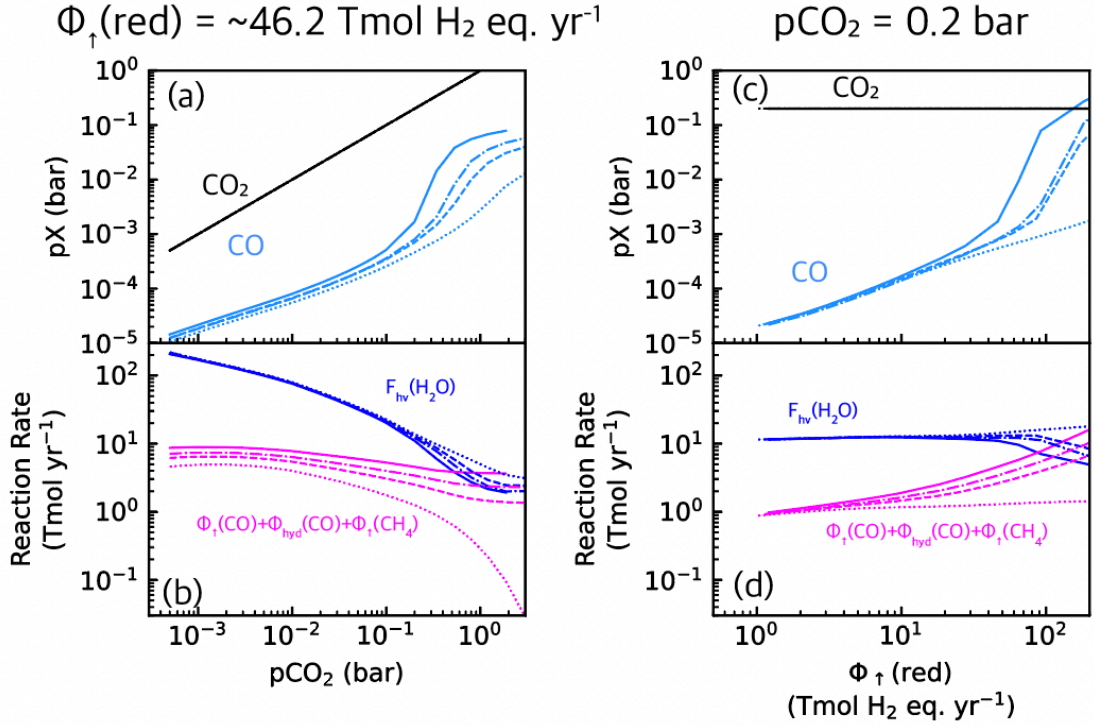

**Figure B1.** Steady-state response of atmospheric $p$CO (blue) level (a, c) and a relationship between the photodissociation rate of $H_2O$ (blue) and input rate of reduced C species (CO and $CH_4$) (b, d) to changes in atmospheric $p$CO$_2$ (a–b) and volcanic outgassing rate of reducing gases (c–d). The solid line is the standard run. The dashed line represents the result without volcanic $CH_4$ outgassing. The dot-dashed line is the result without volcanic CO outgassing. The dashed line represents the result without volcanic CO and $CH_4$ outgassing.

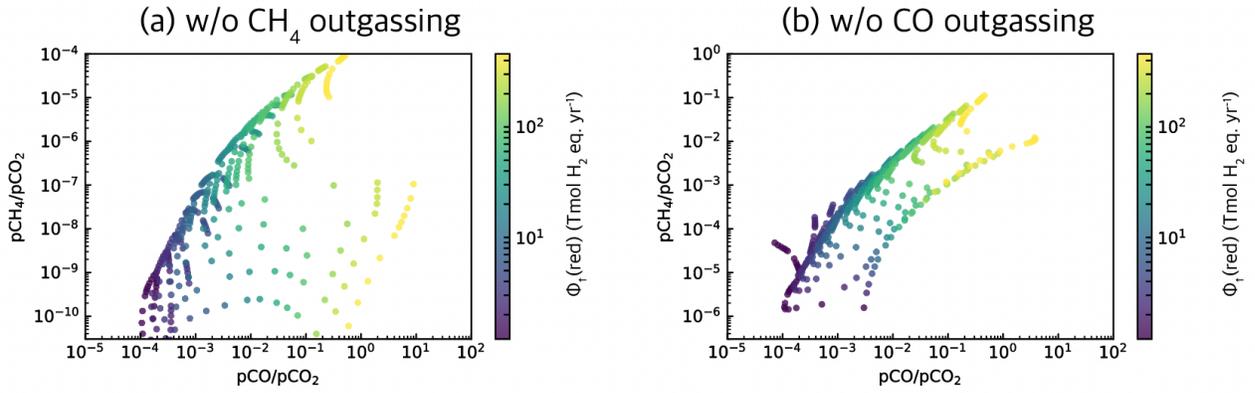

**Figure B2.** Steady-state response of atmospheric CH$_4$/CO$_2$ and CO/CO$_2$ for different types of central stars. (a) Young Sun (4 Ga), (b) present Sun (G2V), (c) σ Bootis (F2V), and (d) ε Eridani (K2V). Colors represent the volcanic outgassing flux of reducing gases, Φ$_\uparrow$(red).

APPENDIX C

**Response of atmospheric chemistry under CO-runaway conditions**

The steady-state response of vertical profiles of the atmospheric $CO_2$, OH, CO, $H_2$, and $CH_4$ levels with respect to different values of atmospheric $pCO_2$ levels are shown in Figure C1. As the atmospheric $pCO_2$ increases, the photodissociation rate of $CO_2$ also increases, especially at altitudes of 10–50 km (Figure C1f):

$$CO_2 + h\nu \rightarrow CO + O(^3P). \qquad (R8)$$

There is another peak at altitudes of $> \sim 50$ km owing to the following pathway:

$$CO_2 + h\nu \rightarrow CO + O(^1D). \qquad (R9)$$

The rate of $H_2O$ photodissociation decreases with altitude in the troposphere (Figure C1g) because the wavelength of $H_2O$ photodissociation ($< \sim 240$ nm) and $CO_2$ dissociation ($< \sim 204$ nm; peaked at $\sim 140$ nm) broadly overlaps:

$$H_2O + h\nu \rightarrow H + OH. \qquad (R10)$$

This reaction plays a key role as an ultimate source of OH radicals in the atmosphere. The atmospheric OH is one of the primary sinks of atmospheric CO and $CH_4$:

$$CO + OH \rightarrow CO_2 + H, \qquad (R11)$$

$$CH_4 + OH \rightarrow CH_3 + H_2O. \qquad (R12)$$

Reaction R11 is a dominant sink of CO in the absence of CO runaway. The concentration of OH radicals tends to be small at altitudes $< \sim 60$ km, especially when the atmospheric CO is high.

When the atmospheric $pCO_2$ increases to approximately 0.3 bar, the mixing ratio of CO, $f$CO, increases dramatically to $> \sim 0.01$, representing the occurrence of CO runaway. Once it happens, Reaction R11 is overwhelmed by the $CO_2$ photodissociation (Reaction R8 and R9) because the $H_2O$ photolysis (i.e., primary OH production) is suppressed owing to the shielding by atmospheric $CO_2$ (Figure C1g). This is the primary mechanism responsible for CO runaway. In CO runaway regimes, the three-body reaction with $O(^3P)$ converts much–but not all–of the excess CO into $CO_2$, (This reaction is slow before CO runaway because it is spin-forbidden):

$$CO + O(^3P) + M \rightarrow CO_2 + M. \qquad (R13)$$

As a result, the $O(^3P)$ mixing ratio in the lower atmosphere becomes very small. Thus, CO runaway is accompanied by the transition of the buffering mechanism of CO in the atmosphere (from Reaction R11 to R13). The surface dry and wet deposition of CO also helps to remove the excess CO from the atmosphere.

Figure 3d also demonstrates that higher input rates of the reducing gases ($\Phi_1(red)$) result in lower threshold values of atmospheric $pCO_2$. When the outgassing flux of reducing gas increases, the photodissociation rate of $H_2O$ in the atmosphere decreases. This would be associated with the increase in the atmospheric $pCH_4$:

$$CH_4 + h\nu \rightarrow {}^1CH_2 + H_2 \qquad , \qquad (R14)$$

$$CH_4 + h\nu \rightarrow {}^3CH_2 + H_2 \qquad , \qquad (R15)$$

$$CH_4 + h\nu \rightarrow CH_3 + H \qquad . \qquad (R16)$$

The photodissociation of $CH_4$ occurs at high altitudes (~70–90 km) and it absorbs at the wavelength overlapped with the photodissociation of $H_2O$. Therefore, this works to decrease the photodissociation of $H_2O$ in the troposphere and the OH abundance, which would also help CO runaway.

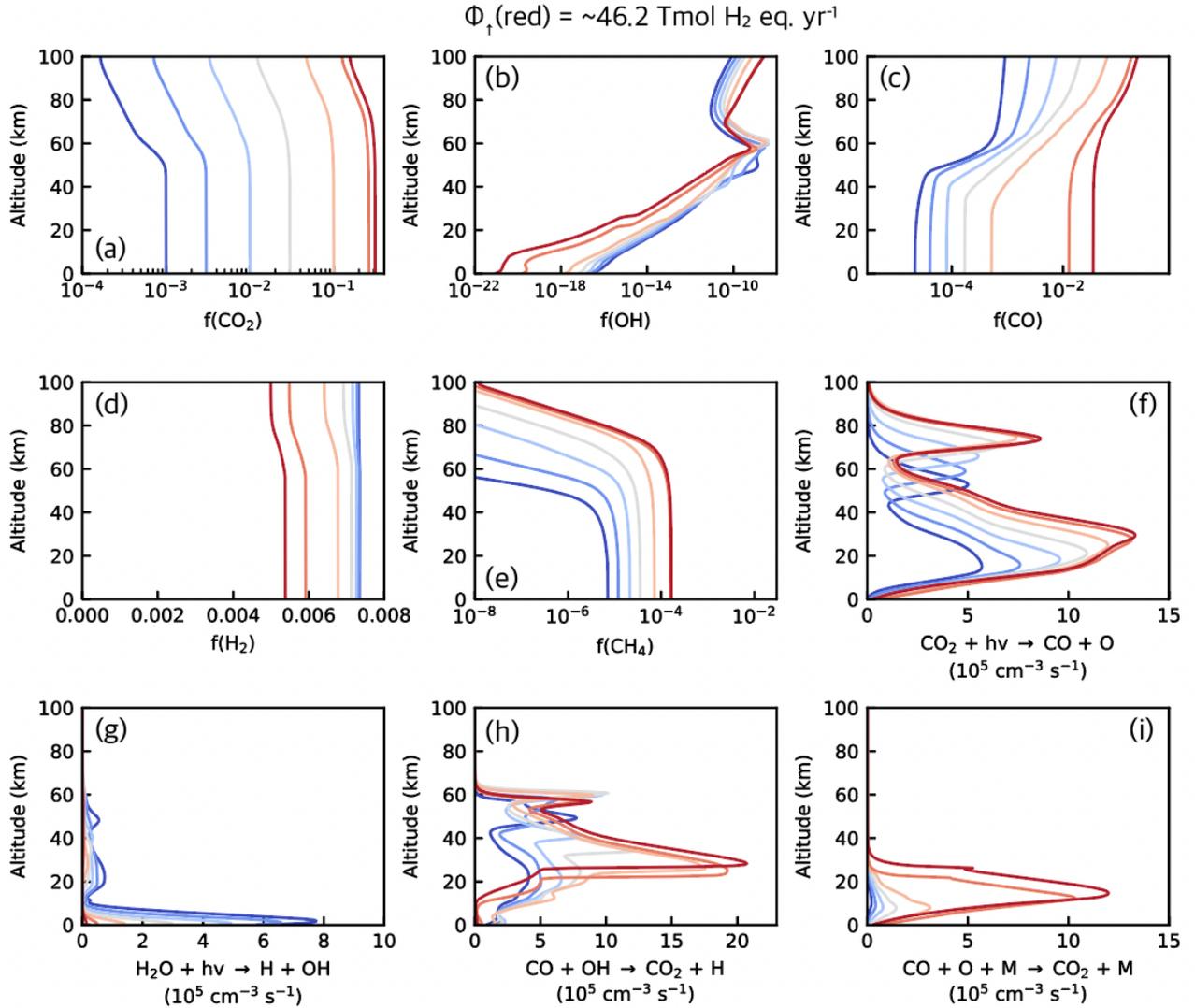

**Figure C1.** Steady-state response of atmospheric chemistry to changes in $p$CO$_2$, assuming the solar spectrum of young Sun (4 Ga). (a) $f$(CO$_2$), (b) $f$(OH), (c) $f$(CO), (d) $f$(H$_2$), (e) $f$(CH$_4$), (f) CO$_2$+hv, (g) H$_2$O+hv, (h) CO+OH, and (i) CO+O+M. The total volcanic outgassing flux of reducing gases, $\Phi_\uparrow$(red), of 46.2 Tmol H$_2$ equiv. yr$^{-1}$ was assumed. Colors represent different $f$(CO$_2$) values.

APPENDIX D

**Chemistry of organic compounds in the prebiotic ocean**

The CO concentrations in the ocean are evaluated by using a simple oceanic chemistry model of Kharecha et al. (2005). In this model, the abiotic chemistry of CO, formate, and acetate in the prebiotic ocean is represented by a simple two-box ocean model. For this calculation, we set the temperature in the ocean to be 273.15 K. The concentrations of CO and formate increases with increasing atmospheric $p$CO (Figure D1). The concentration of CO is slightly higher in the surface ocean than in the deep ocean because it forms formate via a hydration with $OH^-$. The concentration of formate is higher in the deep ocean by more than four orders of magnitudes because formate is converted into acetate in the surface ocean.

The maximum concentration of formaldehyde ($H_2CO$) in the ocean can be estimated by following previous studies (Holland 1984; Harman et al. 2013). Here we assumed $H_2CO$ deposited from the atmosphere to the ocean is decomposed to $H_2$ and CO via reaction R1 once it passes through in the hydrothermal vents. By adopting the present water circulation rate of the mid-ocean ridge hydrothermal vents ($\sim 1.4 \times 10^{14}$ L $yr^{-1}$) and neglecting all other possible loss processes, we can calculate an upper limit on the dissolved $H_2CO$ concentration in the ocean. Under CO runaway conditions, the supply rate of formaldehyde from the atmosphere to the ocean decreases (Figure 5b) because the rate of Reaction R4 decreases in the atmosphere, leading to a decline in $H_2CO$ availability in the troposphere, and hence the oceanic concentration of $H_2CO$. Nevertheless, the maximum concentration of $H_2CO$ could be very high (an order of $\sim 1$ mM) even under the CO-runaway conditions.

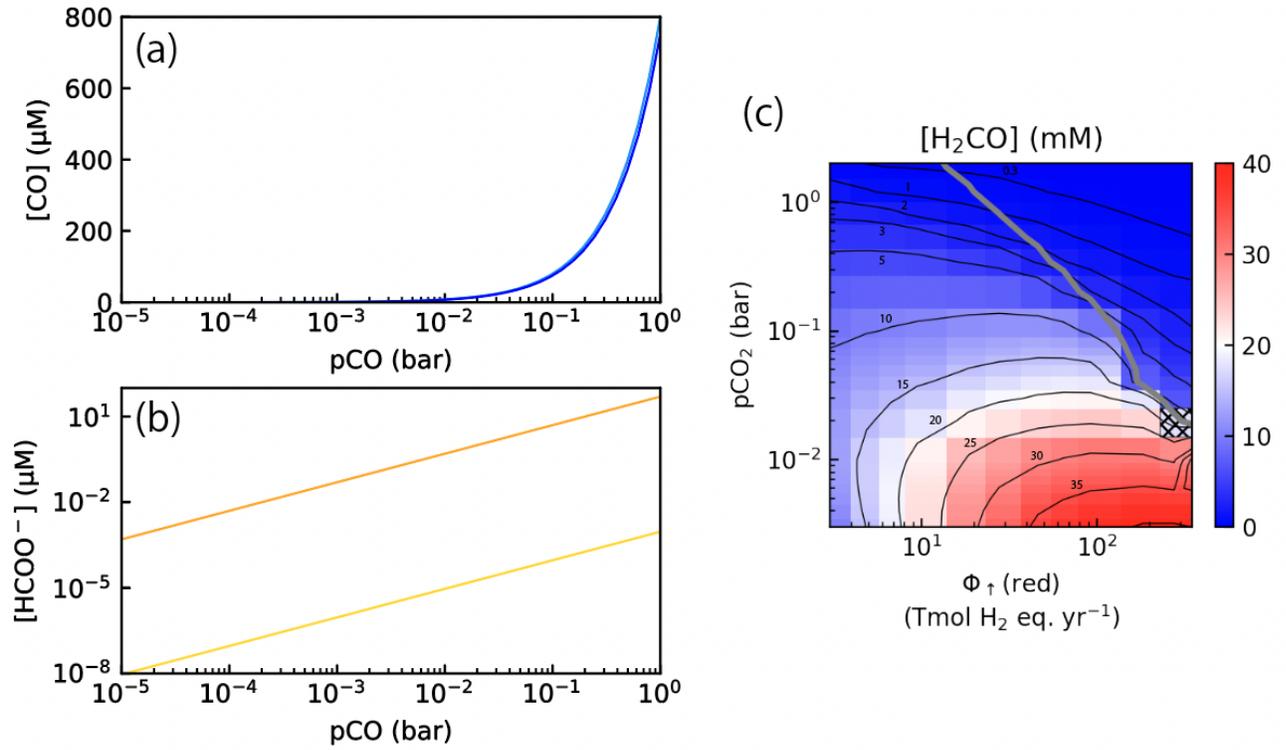

**Figure D1.** The responses of concentrations of (a) CO in the surface and deep oceans (light and dark blue lines, respectively) and (b) HCOO⁻ in the surface and deep oceans (yellow and orange lines, respectively). (c) The upper limit on the dissolved concentration of $H_2CO$ in the ocean as a function of the reduced gas outgassing rate and atmospheric $pCO_2$.